\documentclass[aps,prb,groupedaddress]{revtex4}
\usepackage{amsfonts}
\usepackage{amssymb}
\usepackage{amsmath}
\usepackage{color}
\usepackage{graphicx}

\def\bd#1{\mbox{\boldmath$\displaystyle\mathbf{#1}$} }    
\def\dd{\operatorname{d}}

\begin{document} 
\def\singlespacing{\baselineskip=13pt}	\def\doublespacing{\baselineskip=18pt}
\singlespacing


\title{Effective wave numbers for thermo-viscoelastic media containing random
configurations of spherical scatterers } 

\author{Francine Lupp\'e$^{a}$, Jean-Marc Conoir$^{b}$, Andrew N. Norris$^{c}$}

\affiliation
{$^{a}$Laboratoire Ondes et Milieux Complexes, FRE CNRS 3102, Groupe Ondes Acoustiques,
Universit\'e du Havre, place R. Schuman, 76610 Le Havre, France, \\ 
$^{b}$Institut Jean Le Rond d'Alembert,
UPMC Universit\'e Paris 06, UMR 7190, F-75005 Paris, France, \\ 
$^{c}$Mechanical and Aerospace Engineering, Rutgers University,
Piscataway, NJ 08854, USA.}


\begin{abstract}

The dispersion relation is derived for the coherent waves in fluid or elastic media supporting viscous and thermal effects and containing randomly distributed spherical scatterers. The formula obtained is the generalization of Lloyd and Berry's [Proc. Phys. Soc. Lond. \textbf{91},  678-688, 1067], the latter being limited to fluid host media, and it is the three-dimensional counterpart of that derived by Conoir and Norris [Wave Motion \textbf{47}, 183-197, 2010] for cylindrical scatterers in an elastic host medium.
\end{abstract}

\maketitle

\section{Introduction}\label{sec1}

This study considers wave propagation through a homogeneous isotropic host medium containing a large number of randomly and uniformly located spherical scatterers.  It is well known that for small enough concentrations  the physical medium may be replaced by an effective homogeneous medium in which coherent waves propagate. These waves correspond to the average, over all possible locations of the scatterers, of the multiply-scattered field in the actual host medium.
Our concern  here is to find the effective wavenumbers of the coherent waves at low frequency and low concentration of spheres. This was achieved some time ago by Lloyd and Berry \cite{Lloyd67} for the case of acoustic waves with $P=1$ ($P$: number of waves that propagate in the host medium).  A new and clear derivation of Lloyd and Berry's formula was recently given by Linton and Martin \cite{Linton06} using a procedure   developed earlier for cylindrical scatterers in Ref. \cite{Linton05}.  They considered a uniform concentration of scatterers  satisfying the hole correction of Fikioris and Waterman \cite{Fikioris64}, and found solutions of the implicit dispersion relation as an expansion in terms of the concentration of scatterers, up to order 2, under the low frequency assumption.  Generalization of Linton and Martin's formula for cylinders in a fluid \cite{Linton05} has been given by Conoir and Norris \cite{Conoir10} for cylinders in an elastic solid with $P=2$ (compressional and shear waves propagate in elastic solids).  The present paper is an extension of the results of Linton and Martin \cite{Linton06} to homogeneous isotropic host media supporting viscous and thermal (damped) waves for which $P=3$. Examples of such media include viscous fluids \cite{Epstein53,Allegra72} with compressional and transverse waves ($P=2$),  and thermoelastic solids \cite{Anson93,Challis98,Hasheminejad05} in which compressional, transverse and thermal waves  propagate $(P=3) $. The  results obtained here can serve as a starting point to generalize effective medium theories that  include multiple wave interactions, a broad frequency range, and finite concentration levels, such as the ECAH model (Epstein, Carhart, Allegra and Hawley) \cite{Challis05}.

The analysis begins in section \ref{sec2} with a description  of the scalar potentials that are used to describe the wave propagation in the host medium.  The choice of the potentials depends on conditions of symmetries which are discussed.  Multiple scattering theory is used  in section \ref{sec3} to derive the modal equations (Lorentz-Lorenz laws) relating the amplitudes of the $P$-wave types. In section \ref{sec4} the modal equations  are written in  matrix form  which allows us to  obtain a compact form of the dispersion equation  governing the coherent waves. The low concentration and low frequency assumptions are then introduced in section \ref{sec5} and  expansions of the effective wavenumbers  up to second order   in concentration are derived. The effective wavenumbers are expressed in terms of series, and, for the faster wave, acoustic wave in viscous fluids and compressional wave in viscoelastic solids, the wave number is given by an integral relation that generalizes the one of Lloyd and Berry \cite{Lloyd67}.

\section{The different types of coherent waves}\label{sec2}

We first define the multiple scattering problem for identical spheres in homogeneous isotropic host media in which three type of waves are present, compressional ($c$), viscous or shear ($s$) and thermal ($th$).   The multiple scattering theory used here follows the lines of that first developed by Fikioris \& Waterman \cite{Fikioris64}  for acoustic waves.  

\subsection{The different types of waves in the host medium}

The dynamic displacement field $\vec{u}$ in the host medium may always be decomposed as $\vec u=\vec{\nabla}\psi_L +\vec{\nabla}\times \vec{\psi}_R$, with $\psi_L$ the scalar potential for longitudinal (compressional or thermal) wave motion, and $\vec{\psi}_R$ the vector potential associated with rotational waves \cite{Challis98}. Each wave type propagates with its own complex wavenumber $k_{p}$, with $p$ an integer numbering the wave ($1\leq p \leq P$ ). The rotational wave field itself  can generally be partitioned into a shear and a transverse wave, 
corresponding  to the decomposition of the vector potential $\vec{\psi}_R$ as the sum of two orthogonal vectors $\vec{\psi}_R  =\vec{\nabla}\times(\psi r \widehat{e_{r}})+(1/k_{p})\, \vec{\nabla}\times \vec{\nabla}\times(\chi r \widehat{e_{r}})$ with $\widehat{e_{r}}$ the unit vector along the radial coordinates \cite{Brill80}. The Debye potentials $\psi$ and $\chi$ are  associated with the shear ($s$) and transverse ($T$) waves, respectively.  As a consequence, as many as four distinct wave types  can propagate in the host medium ($p=c,th,s,T $).

Our objective  here is to determine the dispersion relations of the coherent waves in the presence of spheres that are randomly and uniformly located in the host medium. \textit{A priori}, there should be one coherent wave for each longitudinal wave, and one for each shear and each transverse wave. This is a natural hypothesis 
if the scatterers are scarce, that is of low volumetric concentration. It is important to note that the dispersion equations are a characteristic of the effective medium itself, and, as such, they should not depend on the type of the incident plane wave, which will be supposed from now on to be a compressional wave. This choice implies symmetries on the scattering by one scatterer \cite{Challis98,Hasheminejad05}, which, in turn, have implications  on the form of the  analytical expressions for the potentials associated with the coherent waves   propagating in the effective medium.

\subsection{Symmetries}

The scatterers are assumed to be randomly and uniformly located in the  semi-infinite region $z>0$ of the host medium, with an harmonic compressional plane wave   incident on the  boundary $z=0$.  Coherent wavenumbers do not depend on the angle of incidence of the incident wave \cite{Linton06}, and  so  with no loss in generality we assume normal incidence.  Let us  first consider  scattering by a single sphere of an incident  compressional plane wave propagating in the $z$ direction. Due to the symmetry of the sphere, the scattered fields do not depend on the azimuthal angle $\varphi$, with spherical coordinates  defined by  the sphere center ($(x,y,z)=r(\sin \theta \cos\varphi,\sin \theta \sin \varphi,\cos\theta)$)  \cite{Challis98,Hasheminejad05}. Consequently, the scattered transverse $T$ wave with pure azimuthal displacement  is zero. The question  now  is whether a transverse coherent wave can arise from the multiple scattering process. The answer is not straightforward, because the independence of a field with respect to the azimuthal angle in a given spherical coordinate system does not guarantee  independence of that same field with respect to the azimuthal angle in another spherical coordinate system. This means that each multiply scattered field is a combination of longitudinal, shear and transverse waves. However, if the scatterers are uniformly distributed, the average introduces an homogenization of the different fields in the medium, and the effective fields exciting a given sphere should not depend on the azimuthal angle $\varphi$, and, thus, the coherent transverse $T$ field should vanish.

Another way of considering the symmetry is to bear in mind  that coherent waves  result mainly from constructive interferences of waves traveling from one sphere to another along a ``straight line'' (the direction of propagation of the coherent waves). The effective coherent fields  maintain the same symmetry as that for the field scattered  by a single sphere.  We therefore conclude that the coherent fields do not depend on the azimuthal direction $\varphi$ in the $(x,y)$ plane perpendicular to the direction of the incident plane wave.  Mathematically, this result is exact up to order 2 in concentration, as shown in Appendix A. In the following, the longitudinal fields are described by scalar potentials of displacement, and the shear fields by the only non-zero component (azimuthal component) of the vectorial potential of displacement related to shear waves. These scalar quantities are denoted by $\varphi^{(p)}$ , with $p$ denoting again the type of the wave ($p=c,s,th$).

\section{Multiple scattering equations}\label{sec3}

Harmonic wave motion is considered with   time dependence exp$(-i\omega t)$ understood.  The notation $c=1$, $s=2$ and $th=3$ is employed, similar to that used in  Ref. \cite{Conoir10}. The potential function $\varphi_S^{(p)} (\vec{r};\vec{r}_{j})$ represents the wave of type $p$ scattered by a target  centered at $\vec{r}_{j}$ and observed at $\vec{r}$; the potential  $\varphi_E^{(p)}(\vec{r};\vec{r}_{j})$ denotes the field of type $p$ that   excites  a scatterer centered at $\vec{r}_{j}$ and observed at $\vec{r}$.   
The following fundamental identity, which is  a straightforward generalization of Eqs.\ (6) from Ref. \cite{Conoir10},  provides the integral equation  governing the coherent fields, denoted by  brackets, in the presence of a uniform and random array of identical scatterers: 
\begin{equation}\label{lup1}
\langle \varphi^{(p)}_E (\vec{r};\vec{r}_1)\rangle =
 \ \delta_{p1} \varphi_{inc}^{(1)} (\vec{r}) + \sum_{q=1}^P \int\dd 
 \vec{r}_j \, n(\vec{r_{j}}, \vec{r}_1) T^{qp}(\vec{r}_j) 
\langle \varphi^{(q)}_E (\vec{r};\vec{r}_j)\rangle .
\end{equation}
The integration in  Eq.\  \eqref{lup1} is  over the semi-infinite region $(z>0)$  containing the spherical scatterers.  The function  $n(\vec{r_{j}},\vec{r}_1)$ is the conditional number density of spheres at $\vec{r}_{j}$ if one is known to be at $\vec{r}_{1}$, see  Ref. \cite{Conoir10}.   In the following we assume a constant density $n_{0}$ of scatterers of radius $a$, and conditional number density given by   the \textit{hole correction}
\begin{equation}\label{lup2}
n(\vec{r}, \vec{r}_j)= 
\begin{cases}
 n_{0}& \text{for}\quad \left|\vec{r}-\vec{r}_j\right|>b , \quad
 \text{with }\ \ b>2a.
\\
0& \text{otherwise}.
\end{cases} 
\end{equation}
The incident compressional wave is assumed to be a damped plane wave
\begin{equation}\label{lup3}
\varphi_{inc}^{(1)}(\vec{r})=e^{ik_{p}z}\,\,\,\text{with}\,\,\, k_{p}=k_{p}^{'}+ik_{p}^{''}\,\,\, (k_{p}^{''}>0).
\end{equation}
The  potential functions  are expressed as infinite series of spherical harmonics that respect the underlying symmetry.  For the reasons described before they do not depend on the azimuthal direction $\varphi$, and so 
\begin{equation}\label{lup4}
\langle \varphi^{(p)}_E (\vec{r},\vec{r}_j)\rangle 
 = \sum_{n=0}^{+\infty} A_n^{(p)} (\vec{r}_j)\, 
 j_n(k_{p}\rho_j ) P_{n} \big( \cos\theta (\vec{\rho}_j) \big)
 \quad \text{with} \ \   \vec{\rho}_j\equiv \vec{r}-\vec{r}_j . 
\end{equation}
It proves useful to introduce   shorthand notation for spherical harmonics  (not to be confused with ordinary Bessel and Hankel function notation) 
\begin{equation}\label{lup5}
J_n(k_{p},\vec{\rho}_j)=
 j_n(k_{p}\rho_j) P_{n} \big( \cos\theta (\vec{\rho}_j) \big), \,\,\,\,\,\,
H_n(k_{p},\vec{\rho}_j)=
 h_n^{(1)}(k_{p}\rho_j) P_{n} \big( \cos\theta (\vec{\rho}_j) \big),
\end{equation}
plus the following definition for the action of the transition operators $T^{qp}$ on a spherical harmonic \cite{Challis98,Hasheminejad05} 
\begin{equation}\label{lup6}
T^{qp}(\vec{r}_j) \, J_n(k_{p},\vec{\rho}_j)=T^{qp}_n \, H_n(k_{p},\vec{\rho}_j).
\end{equation}
The modal coefficients $T^{qp}_n$ in Eq.\  \eqref{lup6} can be  numerically calculated following the procedures of Refs. \cite{Challis98,Hasheminejad05}.
Equation \eqref{lup1} can now  be expressed, using Eqs. \eqref{lup3} and \eqref{lup4}, 
\begin{equation}\label{lup7}
\sum_{n=0}^{+\infty} A_n^{(p)} (\vec{r}_1) J_n(k_{p},\vec{\rho}_1)=
\ \delta_{p1} e^{ik_{p}z_{1}} +  \sum_{n=0}^{+\infty} \sum_{q=1}^P \int\dd 
\vec{r}_j \, n(\vec{r_{j}}, \vec{r}_1) A_n^{(q)} (\vec{r}_j) T^{qp}_n H_n(k_{p},\vec{\rho}_j).
\end{equation}
Decomposition of the damped incident plane wave into spherical harmonics in the coordinate system centered at $\vec{r}_1$ leads to
\begin{align}\label{lup8}
&\sum_{n=0}^{+\infty} \big[ A_n^{(p)} (\vec{r}_1)-\delta_{p1}i^{n}(2n+1)e^{ik_{p}z_{1}} \big] J_n(k_{p},\vec{\rho}_1) 
\nonumber \\
& \qquad  \qquad \qquad  \qquad  
=\sum_{n=0}^{+\infty}  \sum_{q=1}^P  \int\dd 
\vec{r}_j \, n(\vec{r_{j}}, \vec{r}_1) A_n^{(q)} (\vec{r}_j) T^{qp}_n H_n(k_{p},\vec{\rho}_j).
\end{align}
The  addition theorem \cite{Cruzan62} allows us to write the series in the right hand side of Eq.\  \eqref{lup8} as a function of the coordinates centered on $\vec{r}_1$,  
\begin{align}\label{lup9}
H_n(k_{p},\vec{\rho}_j)
&=\sum_{\nu=0}^{+\infty} \sum_{\mu=-\nu}^{+\nu} \sum_{\ell =0}^{+\infty} 
(-1)^{\mu}\, i^{\nu+\ell-n}(2\nu+1)G(0,n;-\mu,\nu;\ell)e^{i\mu\varphi(\vec{\rho}_1)}e^{-i\mu\varphi(\vec{r_{1j}})}
\nonumber \\
&
\times h_n^{(1)}(k_{p}r_{1j})P_{\ell}^{-\mu} \big( \cos\theta (\vec{r_{1j}}) \big)
j_\nu(k_{p}\rho_{1})P_{n}^{\mu} \big( \cos\theta (\vec{\rho}_1) \big) \quad \text{with}  
\quad \vec{r_{1j}}=\vec{r_{1}}-\vec{r_{j}},
\end{align}
where the Gaunt coefficients $G(0,n;-\mu,\nu;m)$ are defined by  
\begin{equation}\label{lup10}
P_{n}^{m}\big(\cos\theta\big)P_{\nu}^{\mu}\big(\cos\theta\big)=\sum_{\ell=0}^{\infty} G(m,n;\mu,\nu;\ell)
P_{\ell}^{m+\mu}\big(\cos\theta\big).
\end{equation}

By assumption, the incident damped plane wave impinges on the ($z=0$) interface at normal incidence, and  gives rise to damped coherent waves  that propagate and are attenuated in the same direction $z$.   Accordingly, we seek coherent plane  wave solutions   obeying the Snell-Descartes laws of refraction.
 We thus search for the solutions of Eq.\  \eqref{lup8} in the form
\begin{equation}\label{lup11}
A_n^{(q)} (\vec{r}_j)=i^{n}(2n+1)\sum_{p=1}^{P} A_n^{(qp)} e^{i\xi_{p}z_{j}}. 
\end{equation}
Substituting from Eqs.\ \eqref{lup9} and  \eqref{lup11} into Eq.\ \ \eqref{lup8}, making the change of variables of integration $\vec{r_{j}}=\vec{r_{1}}-\vec{r_{1j}}$, and noting that the integration over $d\varphi_{1j}=d\varphi(\vec{r_{1j}})$ with $d\vec{r_{1j}} =r_{1j}^{2}\sin \theta_{1j}dr_{1j}d\theta_{1j}d\varphi_{1j}$ gives rise to zero except for $\mu=0$ because of the term exp$\big[-i\mu \varphi(\vec{r_{1j}})\big] $, yields
\begin{equation}\label{lup12}
A_n^{(pk)}e^{i\xi_{k}z_{1}}=\delta_{p1}e^{ik_{p}z_{1}}+
\sum_{q=1}^{P} \sum_{\nu=0}^{+\infty} \sum_{\ell =0}^{+\infty} i^{\ell}(2\nu+1)T_{\nu}^{qp} A_{\nu}^{(qk)}
G(0,\nu;0,n;\ell) I_{\ell}^{(p)}(\xi_{k}),
\end{equation}
where the quantity $I_{\ell}^{(p)}$ is 
\begin{equation}\label{lup13}
I_{\ell}^{(p)}(\xi_{k})= \int\dd 
\vec{r}_j \, n(\vec{r_{j}}, \vec{r}_1) e^{i\xi_{k}z_{j}} h_{\ell}^{(1)}(k_{p}r_{1j})P_{\ell}\big( \cos\theta (\vec{r_{1j}}) \big).
\end{equation}
Taking into account the hole correction \eqref{lup2}, and following the analysis of  Ref.\  \cite{Fikioris64}, we obtain 
\begin{subequations}\label{lup14}
\begin{align}\label{14a}
I_{\ell}^{(p)}(\xi_{k})&=\frac{2 n_{0}\pi i^{\ell}}{\xi_{k}-k_{p}}
\bigg[ \frac{2b}{\xi_{k}+k_{p}}N_{\ell}^{(p)} (\xi_{k}) e^{i\xi_{k}z_{1}}+\frac{i}{k_{p}^2}e^{ik_{p}z_{1}}\bigg]
\ \ 
\text{with}
\\
N_{\ell}^{(p)} (\xi_{k})&=\xi_{k}bj_{\ell}^{'}(\xi_{k}b)h_{\ell}^{(1)}(k_{p}b)-
k_{p}bj_{\ell}(\xi_{k}b)h_{\ell}^{(1)'}(k_{p}b).
\end{align}
\end{subequations}
Inserting Eq.\  \eqref{lup14}  into Eq.\  \eqref{lup12}, and equating the coefficients of exp$(i\xi_{k}z_{1})$ to zero gives what is known as the Lorentz-Lorenz law (equating the coefficients of exp$(ik_{p}z_{1})$ to zero gives the extinction theorem). Finally, we obtain the  equations for the amplitudes 
$A_n^{(pk)}$,
\begin{equation}\label{lup16}
A_n^{(pk)}=\frac{4 n_{0}\pi b}{\xi_{k}^{2}-k_{p}^{2}}\sum_{q=1}^{P} \sum_{\nu=0}^{+\infty} \sum_{\ell =0}^{+\infty} (-1)^{\ell}(2\nu+1)T_{\nu}^{qp} A_{\nu}^{(qk)} N_{\ell}^{(p)}(\xi_{k}) G(0,\nu;0,n;\ell)
\end{equation}
with $1\leqslant p \leqslant P$ and $1\leqslant k \leqslant P$.

\section{Matrix form of the modal equation}\label{sec4}

We now seek to write Eq.\  \eqref{lup16} in   matrix form in the same way as in Ref.\  \cite{Conoir10,Norris11}.  With no loss of generality,  consider a given effective wave number $\xi_{k}$ ($k=c,s,th$). Equation \eqref{lup16} can therefore be written in   simplified form by dropping the index $k$ ($\xi_{k}=\xi$). Define the non-dimensional parameters 
\begin{equation}\label{lup17}
y_{p}=\xi^{2}-k_{p}^{2}\,\,\,\,\,\,\text{and}\,\,\,\ \varepsilon=-4in_{0}\,,
\end{equation}
and the use of the relation \cite{Abramowitz74}
\begin{equation}\label{lup18}
(-1)^{\ell}G(0,\nu;0,n;\ell)=(-1)^{\nu+n}G(0,\nu;0,n;\ell)
\end{equation}
implies ($1\leqslant p \leqslant P$ )
\begin{equation}\label{lup19}
A_n^{(p)}-\frac{i \varepsilon \pi b}{y_{p}}\sum_{q=1}^{P} \sum_{\nu=0}^{+\infty} \sum_{\ell =0}^{+\infty} (-1)^{\nu+n}(2\nu+1)T_{\nu}^{qp} A_{\nu}^{(q)} N_{\ell}^{(p)}(\xi) G(0,\nu;0,n;\ell)=0.
\end{equation}
In order to reach our objective, we define the infinitely long vectors $|e\rangle$, $\langle e|=|e\rangle^{t}$ and  square matrices $\textbf{T}^{qp}$, $\bar{\textbf{Q}}^{(p)}$ from their components, $e_n=(-1)^{n}$ and
\begin{equation}\label{lup20}
T_{n\nu}^{qp}=\delta_{n\nu}(2\nu+1)T_{\nu}^{qp},\,\,\,\,
\bar{Q}_{n\nu}^{(p)}(\xi)=\frac{\pi}{k_{p} y_{p}} 
\big[ \, ik_{p}b \sum_{\ell =0}^{+\infty} N_{\ell}^{(p)}(\xi) G(0,\nu;0,n;\ell)-1 \, 
\big] (|e\rangle \langle e|)_{n\nu}\,\,.
\end{equation}
We also introduce the unknown vectors  $|a_{p}\rangle$ with components $A_{n}^{(p)}$, the vectors $|e_{p}\rangle$  consisting in the combination  of $P$ zero vectors except at the p$^\text{th}$ place where their components are equal to $\sqrt{\pi/k_{p}}\, |e\rangle$, the $|a\rangle$ vector that is a collection of the $|a_{p}\rangle$ vectors, and the  block matrices
\begin{equation}\label{21}
{\bd {I}_{P}}   =  \begin{bmatrix}
 {\bd I} & {\bd 0} &  {\bd 0}
\\
{\bd 0}   &  {\bd I} &  {\bd 0}
\\
{\bd 0}   &  {\bd 0} &  {\bd I}
\end{bmatrix} ,
\qquad
{\bd {T}}   =  \begin{bmatrix}
 {\bd {T}^{11}} & {\bd {T}^{21}} &  {\bd {T}^{31}}
\\
{\bd {T}^{12}} & {\bd {T}^{22}} &  {\bd {T}^{32}}
\\
{\bd {T}^{13}} & {\bd {T}^{23}} &  {\bd {T}^{33}}
\end{bmatrix} ,
\qquad
{\bd {\bar{Q}}}   =  \begin{bmatrix}
 {\bd {\bar{Q}}^{(1)}} & {\bd 0} &  {\bd 0}
\\
{\bd 0}   &  {\bd {\bar{Q}}^{(2)}} &  {\bd 0}
\\
{\bd 0}   &  {\bd 0} &  {\bd {\bar{Q}}^{(3)}}
\end{bmatrix} .
\end{equation}
Using  these definitions  the system of equations \eqref{lup19} can be replaced by the equivalent  condition
\begin{equation}\label{lup22}
\big\{ \bd I_{P}-\varepsilon \bar{\textbf{Q}} \textbf{T} - \varepsilon \sum_{p=1}^{P} \frac{|e_{p}\rangle \langle e_{p}|}{y_{p}} \textbf{T} \big\}\, |a\rangle=|0\rangle.
\end{equation}
The matrices and vectors as defined are based on the assumption $P=3$.  For smaller values of $P$, or equivalently, no coupling between the wave types
$({\bd {T}^{qp}} = {\bd {T}^{qq}} \delta_{qp})$,  Eq.\ \eqref{lup22} decouples into separate equations for each wave type, each analogous to the acoustic $(P=1)$ case.

Equation \eqref{lup22} embodies the multiple scattering of the $P$ wave types in a single consistency relation.  Its  structure has been specifically chosen so that it is the same for spherical and cylindrical geometries \cite{Conoir10} whatever the number of waves ($P=1,2,3$). We can therefore  follow the  procedures as  developed in Ref.\  \cite{Conoir10,Norris11} for the cylindrical case.  After multiplication on the left hand side by $\textbf{T}^{\frac{1}{2}}$ and introduction of the following quantities
\begin{equation}\label{lup23}
|f_{p}\rangle=\textbf{T}^{\frac{1}{2}} |e_{p}\rangle,\,\,\,\,
\langle f_{p}|=\langle e_{p} | \textbf{T}^{\frac{1}{2}},\,\,\,\,
|b \rangle=\textbf{T}^{\frac{1}{2}} |a \rangle
\,\,\,\, \text{and}\,\,\,\
\bd {Q}=\textbf{T}^{\frac{1}{2}} \bd {\bar{Q}} \textbf{T}^{\frac{1}{2}},
\end{equation}
we get
\begin{equation}\label{lup24}
\big\{ \bd I_{P}-\varepsilon (\bd I_{P}-\epsilon \textbf{Q})^{-1} \sum_{p=1}^{P} \frac{|f_{p}\rangle \langle f_{p}|}{y_{p}} \big\}\, |b\rangle=|0\rangle.
\end{equation}
The modal equation is obtained by setting to zero the determinant associated with the infinite homogeneous system of equations Eq.\  \eqref{lup22} \cite{Fikioris64,LeBas05}.    This can be written in a much simpler form by writing  
Eq.\  \eqref{lup24} in the form 
\begin{equation}\label{lup241}
\big\{ \bd I_{P} +\sum_{p=1}^{P} |g_{p}\rangle \langle f_{p}| \big\}\, |b\rangle=|0\rangle,
\quad \text{where}\ \
|g_{p}\rangle \equiv 
-\varepsilon y_p^{-1} (\bd I_{P}-\epsilon \textbf{Q})^{-1}|f_{p}\rangle .
\end{equation}
Noting that the matrix in \eqref{lup241} is the sum of the infinite dimensional  identity  plus a matrix of rank $P$, the determinant can therefore be reduced to one for a $P$-dimensional matrix  through the use of the identity 
\begin{equation}\label{lup242}
\det\bigg( \bd I_{P} +\sum_{p=1}^{P} |g_{p}\rangle \langle f_{p}| \bigg) = 
\det \big( \bd i_{P} + \bd m\big) 
\quad \text{where}\ \
m_{qp} =  \langle g_{q}| f_{p}\rangle ,
\end{equation}
and $\bd i_{P}$ is the identity matrix of dimension $P$.   Setting the determinant in Eq.\  \eqref{lup242} to zero yields
\begin{equation}\label{lup25}
  \begin{vmatrix}
 1 -  \epsilon \frac{M_{11}}{y_1}  & - \epsilon \frac{M_{21}}{y_2} & -  \epsilon \frac{M_{31}}{y_3}
\\ & \\ 
-  \epsilon \frac{M_{12}}{y_1} &  1 -  \epsilon \frac{M_{22}}{y_2} & -  \epsilon \frac{M_{32}}{y_3}
\\ & \\ 
-  \epsilon \frac{M_{13}}{y_1} &  - \epsilon \frac{M_{23}}{y_2} &  1 -  \epsilon \frac{M_{33}}{y_3}
\end{vmatrix}
=0, 
\end{equation}
or equivalently 
\begin{equation}\label{lup27}
\begin{vmatrix}
 y_{1} -  \epsilon M_{11}  & - \epsilon M_{21} & -  \epsilon M_{31}
\\ & \\ 
-  \epsilon M_{12} &  y_{2} -  \epsilon M_{22} & -  \epsilon M_{32}
\\ & \\ 
-  \epsilon M_{13} &  - \epsilon M_{23} &  y_{3} -  \epsilon M_{33}
\end{vmatrix}
=0.
\end{equation}
where the matrix elements are 
\begin{align}\label{lup26}
M_{qp}(\xi) &=\langle f_{q}| (\bd I_{P}-\epsilon \textbf{Q})^{-1}|f_{p}\rangle
\nonumber \\
 &= 
\langle e_{q}| \textbf{T}|e_{p}\rangle+ \sum_{n=1}^{+\infty} \langle e_{q}| \big( \textbf{T} \bar{\textbf{Q}}\textbf{T} \big)^{n}|e_{p}\rangle \varepsilon^{n}
\nonumber \\
&= 
M_{qp}^{(0)}+\sum_{n=1}^{+\infty} M_{qp}^{(n)} \varepsilon^{n}.
\end{align}
This indicates that the elements of the modal equation Eq.\  \eqref{lup27} can be calculated without evaluation of the square root matrix $\textbf{T}^{\frac{1}{2}}$.
It is also evident that the above reduction in the size of the system determinant  is valid as long as the matrix $\bd I_{P}-\epsilon \textbf{Q}$ is non-singular, which is certainly valid for small $\varepsilon$.  
 Equation  \eqref{lup27} is the fundamental equation for determining the coherent wavenumbers $\xi_{p}$.

\section{Asymptotic solutions of the wavenumber equation}\label{sec5}

This section considers asymptotic expansions of the solutions, valid in different limits: first for low concentration and then for low frequency.

\subsection{Low concentration expansion}

Rather than working with the wavenumber directly it is more convenient to expand the solutions of Eq.\ \eqref{lup27} about one of the three leading order solutions $y_p = 0$ ( $\xi=k_{p}$). The non-dimensional parameters $\epsilon$ is small at  low concentration,   $|\epsilon| \ll 1$,  and we therefore  assume  a formal asymptotic expansion in $\epsilon$:  
\begin{equation}\label{lup28}
y_{p}=\varepsilon y_{p}^{(1)}+\varepsilon^{2} y_{p}^{(2)}+...\,\,.
\end{equation}
Inserting the asymptotic expansion into the modal equation \eqref{lup27} provides (\textit{cf}. Appendix B)
\begin{subequations}\label{lup2930}
\begin{align}\label{lup29}
y_{p}^{(1)}&=M_{pp}^{(0)}(k_{p}),
\\
y_{p}^{(2)}&=M_{pp}^{(1)}(k_{p})+\sum_{q \neq p}\frac{M_{pq}^{(0)}(k_{p})M_{qp}^{(0)}(k_{p})}{k_{p}^{2}-k_{q}^{2}}.
\label{lup30}
\end{align}
\end{subequations}
It follows from  Eqs.\ \eqref{lup20} and \eqref{lup26} that
\begin{subequations}\label{lup31}
\begin{align}\label{lup31a} 
M_{qp}^{(0)} &=\frac{\pi}{\sqrt{k_{q} k_{p}}} \sum_{n=0}^{\infty} (2n+1) T_{n}^{qp},
\\
M_{pp}^{(1)}(k_{p}) &=\frac{\pi}{k_{p}}\sum_{q=1}^{3} \sum_{n=0}^{\infty}\sum_{\nu=0}^{\infty} (-1)^{n+\nu}(2n+1)(2\nu+1)
T_{n}^{qp} \bar{Q}_{n \nu}^{(q)}(k_{p})T_{\nu}^{pq},
\label{lup32}
\end{align}
\end{subequations}
with
\begin{subequations}\label{lup33}
\begin{align}\label{lup33a}
\bar{Q}_{n \nu}^{(q)}(k_{p}) =&\frac{i\pi b}{k_{p}^{2}-k_{q}^{2}}(-1)^{n+\nu}
\bigg\{ \frac{i}{k_{q}b}  + \sum_{\ell =0}^{+\infty} G(0,\nu;0,n;\ell)
\nonumber \\
 & 
 \quad  \times \big[ k_{p}b j_{\ell}^{'}(k_{p}b) h_{\ell}^{(1)}(k_{q}b)-
k_{q}b j_{\ell}(k_{p}b) h_{\ell}^{(1)'}(k_{q}b) \big] 
 \bigg\},  \ \
 q\neq p,
\\
\bar{Q}_{n \nu}^{(p)}(k_{p}) =&  -\frac{i \pi b^{2}}{2 k_{p}}(-1)^{n+\nu} 
\sum_{\ell =0}^{+\infty}  G(0,\nu;0,n;\ell)
\bigg\{ 
 j_{\ell}^{'}(k_{p}b) \big( h_{\ell}^{(1)}(k_{p}b)
 + k_{p}b  h_{\ell}^{(1)'}(k_{q}b)  \big) 
\nonumber \\
 & \qquad
+\frac{1}{k_{p}b} \big[ (k_{p}b)^{2}-\ell(\ell+1) \big] j_{\ell}(k_{p}b) h_{\ell}^{(1)}(k_{p}b)
\bigg\}
\label{lup34}
\end{align}
\end{subequations}
where  the following relation has been  taken into account,
\begin{equation}\label{lup35}
\sum_{\ell =0}^{+\infty} G(0,\nu;0,n;\ell)=1.
\end{equation}

\subsection{Low frequency expansion}

The long wavelength limit is defined as  $k_{p}b\rightarrow 0$, in which case  Eqs.\ \eqref{lup33}  reduce to  (see Eq.\ \eqref{lup35})
\begin{subequations}\label{lup36}
\begin{align}\label{lup36a}
\bar{Q}_{n \nu}^{(q)}(k_{p})
&=\frac{\pi}{k_{q}(k_{p}^{2}-k_{q}^{2})}(-1)^{n+\nu}
\sum_{\ell =0}^{+\infty}  G(0,\nu;0,n;\ell)\, 
\big[   \big( \frac{k_{p}}{k_{q}} \big)^{\ell} -1  \big], \ \ q\ne p ,
\\
\bar{Q}_{n \nu}^{(p)}(k_{p}) &= \frac{\pi}{2 k_{p}^{3}}(-1)^{n+\nu} \sum_{\ell =0}^{+\infty} \ell \,
G(0,\nu;0,n;\ell).
\label{lup37}
\end{align}
\end{subequations}
The low frequency version of  Eq.\ \eqref{lup32} is
\begin{align}\label{lup38}
M_{pp}^{(1)}(k_{p}) &=\frac{\pi^{2}}{2k_{p}^{4}}
\sum_{n=0}^{+\infty} \sum_{\nu=0}^{+\infty} \sum_{\ell =0}^{+\infty} (2n+1)(2\nu+1)
\, G(0,\nu;0,n;\ell) 
\nonumber \\
 & \quad 
 \times 
\bigg\{
\ell \, T_{n}^{pp}T_{\nu}^{pp} + \sum_{q \neq p}\frac{2 k_p^3}{ k_{q}(k_{p}^{2}-k_{q}^{2})} 
\big[  \big( \frac{k_{p}}{k_{q}} \big)^{\ell} -1
\big] T_{n}^{qp}T_{\nu}^{pq}  \bigg\}. 
\end{align}
and hence  the low frequency expansion of the effective wavenumbers are
\begin{align}\label{lup39}
\frac{\xi_{p}^{2}}{k_{p}^{2}} = &1-4i\pi \frac{n_{0}}{k_{p}^{3}} \sum_{n=0}^{\infty} (2n+1) T_{n}^{pp}
-\frac{8\pi^{2}n_{0}^{2}}{k_{p}^{6}}\sum_{n=0}^{+\infty} \sum_{\nu=0}^{+\infty} \sum_{\ell =0}^{+\infty} (2n+1)(2\nu+1) \, G(0,\nu;0,n;\ell)
\nonumber \\
 & \qquad  \qquad  \qquad 
 \times \bigg\{
\ell \,
T_{n}^{pp}T_{\nu}^{pp}
+\sum_{q \neq p} \frac{2k_{p}^{3}}{ k_{q}(k_{p}^{2}-k_{q}^{2})}  
\big( \frac{k_{p}}{k_{q}} \big)^{\ell} \,  T_{n}^{qp}T_{\nu}^{pq}
\bigg\}.
\end{align}
If the host medium is an ideal fluid, i.e. $P=1$ as in Ref.\ \cite{Linton06}, Eq.\ \eqref{lup39} simplifies to
\begin{align}\label{lup40}
\frac{\xi_{p}^{2}}{k_{p}^{2}} =& 1-4i\pi \frac{n_{0}}{k_{p}^{3}} \sum_{n=0}^{\infty} (2n+1) T_{n}^{pp}
\nonumber \\
 & 
 \quad
-\frac{8\pi^{2}n_{0}^{2}}{k_{p}^{6}}\sum_{n=0}^{+\infty} \sum_{\nu=0}^{+\infty} \sum_{\ell =0}^{+\infty} (2n+1)(2\nu+1)G(0,\nu;0,n;\ell)
\,\ell T_{n}^{pp}T_{\nu}^{pp}  ,
\end{align}
which is exactly the same relation as that obtained by combining Eqs.\ (1.2,4.29, 4.33,C.4) from Ref.\ \cite{Linton06}.  Equation  \eqref{lup39} contains additional terms which are not  in Eq.\ \eqref{lup40}  and   are clearly connected to the coupling between the compressional, shear and thermal waves ($q\neq p$).  These coupling terms are neglected in the ECAH  model (Epstein, Carhart, Allegra and Hawley) \cite{Challis05}.  Equation \eqref{lup39} can therefore serve as  the starting point for further developments, such as the calculation of the Rayleigh limit,  for example, and is the principal result of the paper.

\subsection{Generalization of the Lloyd and Berry formula}

The advantage of the  Lloyd and Berry formula is that it expresses the wavenumber at low concentration in terms of the far-field scattering function only, rather than the T-matrix elements. In the present context, this requires that we  express series in \eqref{lup39} as integrals of the far-field scattering functions defined as follows
\begin{equation}\label{lup41}
f^{qp}(\theta)=\sum_{n=0}^{+\infty}(2n+1)T_{n}^{qp}P_{n}(\cos\theta).
\end{equation}
The analogous acoustic problem involves only the terms in Eq.\ \eqref{lup40} which have been shown to be equivalent to the following expansion in the concentration \cite{Linton06}
\begin{equation}\label{lup42}
\xi_{p}^{2}=k_{p}^{2}+\delta_{1}n_{0}+\delta_{2}n_{0}^{2},
\end{equation}
with
\begin{subequations}\label{lup43}
\begin{align}\label{p43a}
\delta_{1}&=-\frac{4i\pi}{k_{p}}f^{pp}(0),
\\
\delta_{2}&=\frac{4\pi^{2}}{k_{p}^{4}} \bigg\{  \big[f^{pp}(0)\big]^{2} -\big[f^{pp}(\pi)\big]^{2}
+ \int_{0}^{\pi} \frac{\dd\theta}{\sin (\frac{\theta}{2})}\frac{\dd}{\dd\theta} [f^{pp}(\theta)]^{2}    \bigg\},
\label{lup44}
\end{align}
\end{subequations}
where $\delta_{2}$ in Eq.\ \eqref{lup44} is the formula initially given by Lloyd and Berry \cite{Lloyd67}. In order to express the general coupled wave problem in a similar form it is necessary to  represent  the final series in Eq.\ \eqref{lup39} ($q\neq p$) as integrals of the far-field scattering functions.  

The relevant  series that appears in the multi-wave system is 
(see Eq.\ \eqref{lup39})
\begin{equation}\label{lup45}
S(\kappa)=\sum_{n=0}^{+\infty} \sum_{\nu=0}^{+\infty}\sum_{\ell =0}^{+\infty} (2n+1)(2\nu+1)T_{n}^{qp}T_{\nu}^{pq} 
\kappa^{\ell} G(0,\nu;0,n;\ell)
\quad
\text{where} \ \ \kappa=\frac{k_{p}}{k_{q}}.
\end{equation}
Two cases need  to be considered depending upon whether  $|\kappa|<1$ or $|\kappa|>1$.   
For the case in which $|\kappa|<1$, we can define the function
\begin{equation}\label{lup46}
g(\kappa,\theta)=\sum_{m=0}^{+\infty}\kappa^{m}(2m+1)P_{m}(\cos\theta).
\end{equation}
The use of the  following two relations (\textit{cf}. Ref.\ \cite{Abramowitz74})
\begin{subequations}\label{47}
\begin{align}\label{lup47a}
\int_{0}^{\pi} P_{n}(\cos\theta) P_{m}(\cos\theta) \sin \theta \dd\theta &=\frac{2}{2n+1}\,\delta_{nm},
\\
\sum_{n=0}^{+\infty}\kappa^{n}P_{n}(\cos\theta)&=\frac{1}{(1-2\kappa \cos\theta +\kappa^{2})^{\frac{1}{2}}}\,\,\,\,\,\,\, \text{with}\,\,\,\,\,\,\,|\kappa|<1 ,
\end{align}
\end{subequations}
then provides, respectively,
\begin{subequations}\label{48}
\begin{align}\label{lup48}
\kappa^{n}&=\frac{1}{2}\int_{0}^{\pi} g(\kappa,\theta) P_{n}(\cos\theta) \sin \theta \dd\theta,
\\
g(\kappa,\theta)
&=\big[1+ 2\kappa \frac{d}{d\kappa} \big] \frac{1}{(1-2\kappa \cos\theta +\kappa^{2})^{\frac{1}{2}}} \, .
\label{lup50}
\end{align}
\end{subequations}
The product of the far-field functions follows from the definition \eqref{lup41} as 
\begin{equation}\label{lup51}
f^{qp}(\theta)f^{pq}(\theta)=\sum_{n=0}^{+\infty} \sum_{\nu=0}^{+\infty}\sum_{\ell =0}^{+\infty} (2n+1)(2\nu+1)T_{n}^{qp}T_{\nu}^{pq} P_{\ell}(\cos\theta) G(0,\nu;0,n;\ell) .
\end{equation}
Combing the above results and performing the differentiation  in \eqref{lup50} gives 
\begin{equation}\label{lup52}
S(\kappa)=\frac{1}{2} (1- \kappa^2) \int_{0}^{\pi}f^{qp}(\theta)f^{pq}(\theta) 
\big( 1-2\kappa \cos\theta +\kappa^{2} \big)^{-\frac 32}
 \sin \theta \, \dd\theta.
\end{equation}
The series $S(\kappa)$ is therefore convergent since it can be expressed as the  integration of a continuous function over the compact interval [$0,\pi$]. Although we did not manage to express the remaining series in Eq.\ \eqref{lup39} in terms of the form functions when $|\kappa|>1$, the series are always convergent because there is only a limited number of significant values of the $T_{n}^{qp}$ coefficients, see the discussion on this point in Ref.\ \cite{Conoir10}. It is also easy to show that the Gaunt coefficients decrease much more quickly that the function $|\kappa|^{\ell}$ increases with $\ell \rightarrow +\infty$ even for $|\kappa|>1$.
\

If $p$ indicates the faster wave  propagating in the medium, then it is  always true that  $|\kappa|=|k_{p}/k_{q}|<1$ whatever the value of $q\neq p$. This is the case for acoustic waves in viscous fluids and compressional waves in viscoelastic solids.
So, in such cases, if $\xi_{p}$ is the wavenumber of the acoustic or compressional wave, it follows in a straightforward manner from Eqs.\ \eqref{lup39} to \eqref{lup52} that
\begin{equation}\label{lup53}
\xi_{p}^{2}=k_{p}^{2}+\delta_{1}n_{0}+\delta_{2}n_{0}^{2}+\delta_{2}^{(c)}n_{0}^{2}
\end{equation}
with
\begin{align}\label{lup54}
\delta_{2}^{(c)}
&=
\sum_{q \neq p} \frac{16 \pi^{2} }{k_{p} k_{q}(k_{q}^{2}-k_{p}^{2})} \,
S( \frac{k_{p}}{k_{q}})
\notag
 \\
& = \frac{8 \pi^{2} }{k_{p}}\,
\sum_{q \neq p}
\,
\int_{0}^{\pi} \frac{f^{qp}(\theta)f^{pq}(\theta) \, 
 \sin \theta \, \dd\theta }{
\big(k_{p}^2+ k_{q}^2 -2k_{p} k_{q} \cos\theta \big)^{\frac 32}}
 .
\end{align}
The additional term $\delta_{2}^{(c)}$ indicates the coupling between the compressional ($p$), shear and thermal waves ($q\neq p$). The formula given by Eqs.\ \eqref{lup53} combined with \eqref{lup54} generalizes the identity for acoustic waves derived by Lloyd and Berry  \cite{Lloyd67} for which $\delta_{2}^{(c)}=0$. It is of interest to compare the structure of $\delta_{2}^{(c)}$  with that of the corresponding term for the effective quasi-longitudinal  wave in the presence of  cylinders in an elastic host medium \cite[Theorem 1]{Conoir10}. The coefficient for  the latter case is obtained from \eqref{lup54} by removing $\sin \theta $ in the numerator and replacing the power $\frac 32$ in the denominator with $1$, that is, by making the obvious changes one would expect for 2D as compared with 3D.

\appendix

\section{Azimuthal waves at second order}

The aim of this appendix is to show that the coherent fields do not depend on the azimuthal direction $\varphi$ up to second order  in concentration. We refer here to equations in the paper of Linton and Martin \cite{Linton06}. The fact that  it  only considers compressional or acoustic waves ($P=1$)  is not important for the present  demonstration, which is the same for one or for several waves ($P=3$).  We begin with the modal equation for  oblique incidence, (\textit{cf}. \cite[Eq.\ 4.20]{Linton06})
\begin{equation}\label{lupA1}
F_n^{m}+\frac{i n_{0}(4\pi)^{2}(-1)^{m}}{k(k^{2}-K^{2})} \sum_{\nu=0}^{+\infty} \sum_{\mu=-\nu}^{+\nu} \sum_{q=0}^{+\infty} Z_{\nu} F_{\nu}^{\mu}Y_{q}^{\mu-m}(\widehat{\textbf{K}}) \textbf{N}_{q}(Kb) \textbf{G}(n,m;\nu,-\mu;q)=0,
\end{equation}
with the spherical harmonics and the Gaunt coefficients defined by, respectively, 
\begin{subequations}\label{A2}
\begin{align}\label{lupA2}
Y_{n}^{m}(\widehat{\textbf{r}})
=Y_{n}^{m}(\theta,\varphi)
&=(-1)^{m}\sqrt{\frac{2n+1}{4\pi}}\sqrt{\frac{(n-m)!}{(n+m)!}}
P_{n}^{m}(\cos\theta)e^{im\varphi},
\\
Y_{n}^{m}(\widehat{\textbf{r}}) Y_{\nu}^{\mu}(\widehat{\textbf{r}})
&=\sum_{q=0}^{\infty} Y_{q}^{m+\mu}(\widehat{\textbf{r}}) \textbf{G}(n,m;\nu,\mu;q).
\label{lupA3bis}
\end{align}
\end{subequations}
Consequently, it is necessary  to prove that the unknown coefficients $F_{n}^{m}$ are zero except for $m=0$, at least up to order 2 in concentration.  We use the following result for the expansion of $F_{n}^{m}$ up to and including the second order in concentration 
(\textit{cf}. \cite[Eq.\ 4.20]{Linton06} and its preceding equation)
\begin{equation}\label{lupA3}
F_n^{m}= \overline{Y}_{n}^{m}(\widehat{\textbf{K}}) \tilde{F}+n_{0}
\big\{
\overline{Y}_{n}^{m}(\widehat{\textbf{K}}) V
+ \frac{(4\pi)^{2}b}{2k^{2}}(-1)^{m}\tilde{F} \sum_{\nu=0}^{+\infty} \sum_{\mu=-\nu}^{+\nu} Z_{\nu} \overline{Y}_{\nu}^{\mu}(\widehat{\textbf{K}}) 
X_{n\nu}^{m\mu} 
\big\}
\end{equation}
with \cite[Eq.\ 4.27]{Linton06}
\begin{equation}\label{lupA4}
X_{n\nu}^{m\mu}=\sum_{q=0}^{+\infty} Y_{q}^{\mu-m}(\widehat{\textbf{K}})\textbf{G}(n,m;\nu,-\mu;q) \, d_{q}(kb),
\end{equation}
where $\tilde{F}$ is a constant and where $V$ \cite[Eq.\ 4.31]{Linton06} and $d_{q}(kb)$ \cite[Eq.\ 4.23]{Linton06} are functions which do not depend on the indices $n$ and $m$.  Since the coherent wavenumbers do not depend upon the angle of incidence \cite{Linton06}, we may consider the case of  normal incidence for which  $\widehat{\textbf{K}}=(0,0,1)$ \cite[Eq.\ 3.15]{Linton06}, so that
\begin{equation}\label{lupA5}
Y_{q}^{\mu-m}(\widehat{\textbf{K}})=\sqrt{\frac{2q+1}{4\pi}}\delta_{\mu m}.
\end{equation}
Equation \eqref{lupA3} therefore reduces to
\begin{align}\label{lupA6}
F_n^{m} =& \sqrt{\frac{2n+1}{4\pi}}\delta_{m0} \,
\big(\tilde{F}+  n_{0} V \big) +  n_{0} \tilde{F} 
\frac{4\pi b}{2k^{2}}(-1)^{m}  
\nonumber \\
 & \times
\sum_{\nu=0}^{+\infty} \sum_{\mu=-\nu}^{+\nu} \sum_{q=0}^{+\infty} Z_{\nu} \sqrt{(2\nu+1)(2q+1)} \delta_{\mu 0}\, \delta_{\mu m} \, \textbf{G}(n,0;\nu,0;q) d_{q}(kb),
\end{align}
which proves that $F_{n}^{m}=0$ if $m\neq0$ up to second order   in concentration.

Consequently, at normal incidence and up to order 2 in concentration, Eq.\ \eqref{lupA1} becomes 
\begin{equation}\label{lupA7}
F_n^{0}+\frac{i n_{0}(4\pi)^{2}}{k(k^{2}-K^{2})} \sum_{\nu=0}^{+\infty} \sum_{q=0}^{+\infty} Z_{\nu} F_{\nu}^{0}\sqrt{\frac{2q+1}{4\pi}} \textbf{N}_{q}(Kb) \textbf{G}(n,0;\nu,0;q)=0.
\end{equation}

Our results correspond exactly to those of Ref.\ \cite{Linton06}. Indeed, we can easily show that Eq.\ \eqref{lupA7} and Eq.\ \eqref{lup19} with $P=1$ are exactly the same if we make the following identifications: $k=k_{1}$, $K=\xi_{1}=\xi$, $Z_{\nu}=-T_{\nu}^{11}$, $\textbf{N}_{q}(Kb)=ik_{1}bN_{n}^{(1)}(\xi)$ and
\begin{equation}\label{lupA8}
F_{n}^{0}=(-1)^{n}\sqrt{\frac{2n+1}{4\pi}}A_{n}^{(1)},
\quad 
\textbf{G}(n,0;\nu,0;q)=G(0,\nu;0,n;q) 
\sqrt{ \frac{ (2n+1)(2\nu+1)}{ 4\pi (2q+1)}}.
\end{equation}

\section{Some asymptotic expansions}

With no loss of generality  we  consider the first root $y_{1}$. It follows from Eqs.\ \eqref{lup17}, \eqref{lup28}, and \eqref{lup2930} that the asymptotic expansion of  the 11 element of the matrix in Eq.\  \eqref{lup27} is 
\begin{equation}\label{lupB1}
y_{1}-\varepsilon M_{11}=\big[y_{1}^{(1)}-M_{11}^{(0)}(k_{p}) \big]\varepsilon +
\big[y_{1}^{(2)}-M_{11}^{(1)}(k_{p}) \big]\varepsilon^{2}+...\,\, .
\end{equation}
Hence, to leading order Eq.\  \eqref{lup27} becomes 
\begin{equation}\label{lupB2}
\begin{vmatrix}
 y_{1}^{(1)}-  M_{11}^{(0)}(k_{p})  &  0  & 0
\\ & \\ 
-  M_{12}^{(0)}(k_{p})  &  k_{1}^{2}-k_{2}^{2}  & 0
\\ & \\ 
-  M_{13}^{(0)}(k_{p})  &  0 &  k_{1}^{2}-k_{3}^{2}
\end{vmatrix}
=0,
\end{equation}
which  implies the identity   \eqref{lup29}. Inserting the latter into Eq.\ \eqref{lupB1} and Eq.\  \eqref{lup27} then gives,  at the leading order,
\begin{equation}\label{lupB3}
\begin{vmatrix}
 y_{1}^{(2)}-  M_{11}^{(1)}(k_{p})  &  -M_{21}^{(0)}(k_{p})  & -M_{31}^{(0)}(k_{p})
\\ & \\ 
-  M_{12}^{(0)}(k_{p})  &  k_{1}^{2}-k_{2}^{2}  & 0
\\ & \\ 
-  M_{13}^{(0)}(k_{p})  &  0 &  k_{1}^{2}-k_{3}^{2}
\end{vmatrix}
=0,
\end{equation}
from which    Eq.\  \eqref{lup30} follows.



\end{document}